\documentclass[11pt,a4paper]{article}

\usepackage{amsmath}
\usepackage{amsfonts}
\usepackage{amssymb}
\usepackage{graphicx}
\usepackage{bm}
\usepackage{bbm}
\usepackage{epsfig}
\usepackage{latexsym}
\usepackage{url}
\usepackage{cite}
\usepackage{color}
\usepackage{dsfont}
\usepackage{multirow}
\usepackage{hyperref}

\newtheorem{thm}{Theorem}

\newtheorem{lemma}[thm]{Lemma}

\newtheorem{prop}[thm]{Proposition}
\newtheorem{defn}[thm]{Definition}
\newtheorem{rem}[thm]{Remark}

\def\tr{\hbox{\rm tr}\,}

\def\bra{\langle}
\def\ket{\rangle}

\def\pmx{\begin{pmatrix}}
\def\emx{\end{pmatrix}}
\def\R{\mathbb{R}}
\def\C{\mathbb{C}}

\def\Tr{{\rm Tr}}

\def\be{\begin{equation}}
\def\ee{\end{equation}}

\newcommand{\E}{\mathbb{E}}

\newcommand{\N}{\mathbb{N}}

\newcommand{\conv}{\mathop{\mathrm{conv}}}

\newcommand{\qed}{\hfill $\Box$}

\begin{document}

\title
{Exact values of quantum violations in low-dimensional Bell correlation inequalities}

\author
{Ben Li}

\date{\today}
\date{}
\date{{\small }}

\maketitle
\begin{abstract} 
The famous Clauser-Horne-Shimony-Holt (CHSH) inequality certifies a quantum violation, by a factor $\sqrt{2}$, 
of correlations predicted by the classical view of the world in the simplest possible nontrivial measurement setup
(two systems with two dichotomic measurements each). In such setting, this is the largest possible violation, 
which is known as the \emph{Tsirelson bound}. In this paper we calculate the exact values of quantum violations 
for the other Bell correlation inequalities that appear in the setups involving up to four measurements; 
they are all smaller than $\sqrt{2}$. 
While various authors investigated these inequalities via numerical methods, our approach is analytic. 
We also include tables summarizing facial structure of Bell polytopes in low dimensions. 
\end{abstract}
\thispagestyle{empty}

\section{Introduction}
Ever since the seminal paper by Einstein, Podolsky and Rosen \cite{EPR1935} it has been apparent that quantum theory leads to predictions which are incompatible with the classical understanding of physical reality. Specifically, the probability of outcomes of joint measurements upon a quantum state may not fit into the classical probability scheme; that is, the outcomes of each joint measurements may be correlated in a way contradicting 
limitations imposed by local realism or, mathematically, by local hidden variable models. 

While it is the non-classicality of the quantum world, which usually attracts most attention in this context, 
from the mathematical point of view it is equally striking that -- at least for bipartite systems and 
dichotomic measurements -- 
the discrepancy between classical and quantum correlations can not be arbitrarily large: it can not exceed 
the so-called Grothendieck constant. This is a consequence of the seminal work of Tsirelson \cite{Tsirelson1985} 
and the even more famous Grothendieck inequality \cite{grothendieck} from functional analysis. 
Here we shall concentrate on correlation matrices corresponding to such setups and involving up to four measurements. 
Already Tsirelson noticed that,  for two measurements per site, the maximal possible quantum violations is  $\sqrt{2}$ (the Tsirelson bound). The fact that it can be that large was noticed even earlier by Clauser-Horne-Shimony-Holt \cite{CHSH1969}. 
Quantum violations are certified by the so-called Bell inequalities (defined in the next section). While the exact value 
of the Grothendieck constant is between $1.6769$ and $1.7822$, it is now known -- based on numerical arguments -- that the $\sqrt{2}$  bound is valid for up to five measurements per site,  
see \cite{Ki2017} and its references. 
The novelty of the present work is that we calculate analytically the exact values of quantum violations for the two Bell correlation inequalities that appear in the case of four measurements. 
(For smaller setups there are essentially only CHSH-type inequalities.)  Readers familiar with the subject 
and interested just in those analytical arguments may go directly to Section \ref{section:4by4}. 

 \section{Notation and Background}

In this section we recall the definitions of classical and quantum correlation matrices and summarize the 
relevant results of Tsirelson and Grothendieck.

\begin{defn} \label{def:cc}
	A \( m\times n\) real matrix \((a_{ij})\) is called a classical correlation matrix if there exist random variables \((X_i)_{1\le i\le m}\), \((Y_j)_{q\le j\le n}\) defined on a common probability space, satisfying \( |X_i|\le 1, |Y_j|\le 1\) almost surely, and such that \( a_{ij}=\E X_iY_j\) for all $i,j$. The set of classical correlation matrices is denoted by \( {\bf LC}_{m,n}\).
	\end{defn}
	Let us note here that this notion does not coincide with the concept of correlation from statistics  
but rather with the somewhat less frequently used notion of \emph{cross-covariance} (when, additionally,  $\E X_i=\E Y_j=0$). 

\begin{defn} \label{def:qc}
	A \( m\times n\) real matrix \((a_{ij})\) is called a quantum correlation matrix if there is a quantum state $\rho$ on $\C^{d_1}\otimes \C^{d_2})$ (for some $d_1, d_2 \in \N$), Hermitian operators \((X_i)_{1\le i\le m}\) on \( \C^{d_1}\), \( (Y_j)_{q\le j\le n}\) on \( \C^{d_2}\) satisfying \( \|X_i\|\le 1, \|Y_j\|\le 1\) and such that  \( a_{ij}=\Tr \rho (X_i\otimes Y_j)\) for all $i,j$. The set of quantum correlation matrices is denoted by \( {\bf QC}_{m,n}\).
\end{defn} 
	The physical situation that corresponds to a quantum correlation matrix is when two observers perform 
measurements on a shared quantum state $\rho$, with the first (resp., the second) observer having a choice of $m$ 
(resp., $n$) measurements settings with dichotomic (i.e., binary) outcomes.  By a quantum state we mean a 
trace one positive semi-definite operator;  however, this  will not be important in what follows because 
the sets \({\bf LC}_{m,n}\) and \({\bf QC}_{m,n}\) admit simple geometric descriptions given by the following lemma. 
We refer the reader to Chapter 11 of \cite{ABMB2017} for proofs (the second statement was the main insight of \cite{Tsirelson1985}, 
the first is elementary)  
and for further discussion.

\begin{lemma} \label{descpnLCQC} We have
	 \begin{eqnarray*}& &{\bf LC}_{m,n}=\conv \{(\xi_i\eta_j)_{1\le i\le m, 1\le j\le n};
	 	 \xi_i\in \{1,-1\}, \eta_j\in \{1,-1\}\} \ \hbox{ and } \\	 
	& &{\bf QC}_{m,n}=\{ (\langle x_i,y_j\rangle )_{1\le i\le m, 1\le j\le n};x_i,y_j\in \R^{\min \{m,n\}}, |x_i|\le 1, |y_j|\le1 \},
	\end{eqnarray*}
	where \( |\cdot |\) is the Euclidean norm. 
\end{lemma}

It's obvious from the above characterization that \( {\bf LC}_{m,n}\) is a symmetric convex polytope with \( 2^{m+n-1}\) vertices and it's not hard to see that \( {\bf QC}_{m,n}\) is a symmetric convex compact set. We emphasize that 
this property is specific to our setup; in more general settings it is possible that the set of quantum correlations 
is not closed \cite{slofstra2017,paulsenetal2017}. 

A Bell correlation inequality 
 is a linear functional \( \varphi\) with the property that \( \varphi(A)\le 1\) for any classical correlation matrix \( A\in {\bf LC}_{m,n}\). Hence, if  a matrix \(A\notin {\bf LC}_{m,n}\), then necessarily there exists a Bell correlation inequality \( \varphi\) such that \( \varphi(A)>1\). In this case, we say that the Bell inequality \(\varphi\) is violated by \(A\) and the quantity \( \varphi(A)\) is called the violation.  The quantity \( \max_{A\in {\bf QC}_{m,n}} \varphi(A)\) is well defined since \( \varphi\) is a continuous function on the compact  convex set \({\bf QC}_{m,n}\); it is called the (maximal) quantum violation of \( \varphi\). 
 
 It is readily seen that in order to calculate the maximal possible violation \( \varphi(A)\) for a given $A$ it is enough to 
 consider functionals that define facets (i.e., maximal faces) of \({\bf LC}_{m,n}\). (All other Bell inequalities are convex or positive linear combinations of these.) 
  In other words, we understand Bell inequalities if we know the facial structure of the classical correlation polytope or,
  equivalently,  the set of  vertices of the polytope  dual to \({\bf LC}_{m,n}\). In what follows, 
  we will mostly restrict our attention to such functionals.

\medskip 
The connection between correlation matrices and Grothendieck inequality was first studied by B. Tsirelson \cite{Tsirelson1985}.  This connection is encoded in the following theorem (for more background 
we refer interested readers to \cite{ABMB2017}). 
\begin{thm} 
[Grothendieck-Tsirelson Theorem] 
\label{thm:gro-tsi} 
Given  \(K\ge 1\) and positive integers \(m,n,\) the following two conditions are equivalent:\\
1. We have the inclusion
\be {\bf QC}_{m,n}\subset K \,{\bf LC}_{m,n}.  \label{gro-tsi} \ee
2. For any \(m\times n\) real matrix \((m_{i,j})\) and for any real Hilbert space vectors \(x_i, y_j\) with \( |x_i|\le 1, |y_j|\le 1\) we have 
\be
\sum_{i,j}m_{i,j}\langle x_i, y_j\rangle \le K \max_{\xi_i=\pm 1, \eta_j=\pm 1}\sum_{i,j} m_{i,j}\xi_i\eta_j. \label{groineq}
\ee
Moreover, there exists an absolute constant \(K\ge 1\) such that 1. and 2. hold for any positive integers \(m,n\).
\end{thm}
The inequality (\ref{groineq}) is known as Grothendieck's inequality. The best constant \(K\) that works in all instances of  \eqref{groineq}  (and hence in all instances of \eqref{gro-tsi}) is called the real Grothendieck constant and denoted by \( K_G\). We also denote by \( K_G^{(m,n)}\) the best constant such that (\ref{groineq}) holds for given \(m,n\), and \( K_G^{(n)}=K_G^{(n,n)}\). The main results of this paper concern the constant \( K_G^{(4)}\).

Let us now sketch the argument showing the implication  \emph{2.} $\implies$ \emph{1.}  in Theorem \ref{thm:gro-tsi}, 
i.e., the boundedness of quantum violations in the present setting. 
For fixed \(m,n\), consider an arbitrary \(m\times n\) matrix \(M=(m_{ij})\) and rescale it so that  
 \be \max_{\xi_i=\pm 1, \eta_j=\pm 1}\sum_{i,j} m_{i,j}\xi_i\eta_j = 1 . \ee
Then 
-- by the first part of Lemma \ref{descpnLCQC} -- the functional $\varphi_M(A)= \tr M^TA$ is 
a Bell correlation inequality. 
On the other hand -- by the second part of Lemma \ref{descpnLCQC}  and by \eqref{groineq} -- it follows that 
\[ \max_{A\in {\bf QC}_{m,n}} \varphi_M(A) = \max \sum_{i,j} m_{i,j}\langle x_i, y_j\rangle \le K_G^{(m,n)} \leq K_G.\]
 where the second maximum is taken over all unit vectors \( x_i,y_j\) from a real Hilbert space   
 (which \emph{a posteriori} can be taken to be of dimension not exceeding \( \min\{m,n\}\)). 
 
The argument is easily seen to be reversible.  
If follows in particular that determining the Grothendieck constant \(K_G^{(m,n)}\) 
is equivalent to finding the maximal quantum violation for a finite set of \(m\times n\)  Bell correlation inequalities, namely those corresponding to the facets of the local polytope \( {\bf LC}_{m,n}\). It is worth pointing out that, for large \(m,n\), 
 the facial structure of the local polytope is rather complicated, thus the main difficulty in computing \( K_G^{(m,n)}\) lies then in the classification of facets of the local polytope.  The maximal violation of Bell inequalities corresponding to the facets can subsequently be approximated via SDP programming.

\section{Some elementary observations} 
Let  \(M=(m_{i,j})_{1\le i\le m, 1\le j\le n}\) be a matrix with no zero rows nor columns.  
We are interested in maximizing the quantity from 
the left hand side of \eqref{groineq}: 
\be \label{q-objective}
\sum_{i,j}m_{i,j}\langle x_i, y_j\rangle 
\ee
over any real Hilbert space vectors \(x_i, y_j\) with \( |x_i|\le 1, |y_j|\le 1\).   
(The maximum is attained by compactness since it is clearly enough to consider a space of dimension $m+n$.) 
We have 
\begin{prop} \label{prop:extremalcond}
 If the quantity \eqref{q-objective} is maximized, then: 
 
\noindent  (i)  \( x_i, y_j\) are unit vectors from a space $\mathcal{H}$ with  \(\dim \mathcal{H} \le \min \{m,n\}\) 
 
\noindent (ii)  there exist \( k_i, l_j\in \R_{>0}\) for \( i=1,2,...,m\) and \(j=1,2,..., n\) such that 
\begin{eqnarray}
& &\sum_i m_{i,j} x_i=l_j y_j  \ \hbox{ for all }\  j=1,2,..., n  \  \hbox{and}  \label{extremalcond1} \\
& &\sum_j m_{i,j} y_j=k_i x_i   \ \hbox{ for all }\   i=1,2,...,m.  \label{extremalcond2}
\end{eqnarray} 
If the above holds, the extremal value of the quantity \eqref{q-objective} is $\sum_i k_i= \sum_j l_j$. 
\end{prop}
{\bf Proof.}  
As the objective function \eqref{q-objective}  is an affine function of each of the variables \(x_i, y_j\), 
it follows that the maximum is attained on the boundary, i.e., when $|x_i|=|y_j|=1$. 
However, we need to show that the maximum 
\emph{can not}  be attained if one of the vectors is of norm strictly smaller than $1$.

Let \(x_i, y_j\) be a configuration for which \eqref{q-objective} is maximized. 
Note that 
\[
 \sum_{i,j}m_{i,j}\langle x_i, y_j\rangle=\sum_{j}\Big\langle \sum_i m_{i,j}x_i, y_j\Big\rangle 
\leq \sum_{j} \big| \sum_i m_{i,j}x_i\big|, 
 \]
with equality when, for each $j$, either  (a) $\sum_i m_{i,j}x_i=0$, or 
(b) $y_j$ is a unit vector that is a positive multiple of $\sum_i m_{i,j}x_i$.  
Since the latter condition involves exactly the properties stated in \eqref{extremalcond1} and in the assertion (i) 
of the Proposition, we only need to show that (a)  never happens in the extremal configuration. 

To that end, suppose that (for example) $\sum_i m_{i,1}x_i=0$ and that $m_{1,1}>0$ 
(it is here that we use the ``no zero column'' assumption).   
Pick a unit vector \(u\) such that \( u\perp x_i, u\perp y_j\) for \( i=1,2,...,m, j=1,2,...,n\) (embed the Hilbert space $\mathcal{H}$ into a higher dimensional space if necessary).  Next, let \( y_1'=u,  x_1'=(x_1+tu)/\sqrt{1+t^2}, t>0\) and \( x_i'=x_i, y_j'=y_j\) for \( i=2,...,m, j=2,...,n\).  One checks that 
\[ \sum_{i,j}m_{i,j}\langle x_i', y_j'\rangle -\sum_{i,j}m_{i,j}\langle x_i, y_j\rangle 
=\frac{t}{\sqrt{1+t^2}}+\Big(\frac1{\sqrt{1+t^2}}-1 \Big) \sum_{j\neq 1}m_{1,j}\langle x_1, y_j\rangle,
\] 
which is strictly greater than $0$  when $t$ is positive but  sufficiently small.  
It follows that the original configuration was not extremal. 

The assertion \eqref{extremalcond2} and $|x_i|=1$ are shown similarly by exchanging the roles of \(x_i\) and \( y_j\). 
The bound on \(\dim \mathcal{H}\) follows from the fact that in the extremal configuration the linear spans of 
\((x_i)\) and \( (y_j)\) must coincide. Indeed, suppose for example that some  $x_i \not\in \mathcal{K}$, 
where $\mathcal{K}$ is the linear span of \( (y_j)\).  
Then replacing $x_i$'s by their orthogonal projections onto $\mathcal{K}$ 
leads to a configuration with the same value of the objective function \eqref{q-objective}, 
some of whose elements are of norm strictly smaller than $1$, a contradiction.  \qed

\begin{rem} \label{reduction} 
An byproduct of the above argument is the observation that the value of the maximization problem 
\eqref{q-objective} is the same as either of the two problems given below: 
\be \label{q-objective-mod}
\max_{|x_i|=1} \sum_{j} \big| \sum_i m_{i,j}x_i\big|  \ \ \hbox{ or }\ \   \max_{|y_j|=1}    \sum_{i} \big| \sum_j m_{i,j}y_j\big|  .
\ee
 \end{rem}
\begin{rem} \label{lucky} 
In some cases, Proposition \ref{prop:extremalcond} allows to easily find candidates for extremal configurations 
\((x_i)\) and \( (y_j)\).  Denote  $X = [x_1,\ldots, x_m]$ (a matrix whose columns are $x_1,\ldots, x_m$) 
and similarly $Y = [y_1,\ldots, y_n]$. 
Then the equations \eqref{extremalcond1} and \eqref{extremalcond2} can be written compactly as
\[
XM = YL \  \hbox{ and } \ YM^T=XK, 
\]
where $K,L$ are diagonal matrices with diagonal entries respectively $(k_i)$ and $(l_j)$. 
Now,  if $m=n$ and if the rank of $X$ also equals to $n$, we can eliminate $X,Y$ and obtain  $ML^{-1}M^T=K$. 
Examples when this is useful and when it is not are pointed out in Remark \ref{mixed-luck} at the end of this note. 
 \end{rem}

\section{$(2,2) $ Bipartite system and $(3,3) $ Bipartite system}

In the presenting section we consider first the classical polytope \( LC_{2,2}\). We identify \( {\bf LC}_{2,2}\) with a subset of \( \R^4\).  By Lemma \ref{descpnLCQC}, we see that the 8 vertices of  \( {\bf LC}_{2,2}\) are 
\[v_1 = \begin{bmatrix}
1 & 1           \\[0.3em]
1 & 1           \\[0.3em]
          
\end{bmatrix},\ \ \
v_2 = \begin{bmatrix}
-1 & -1           \\[0.3em]
-1 & -1            \\[0.3em]
       
\end{bmatrix};\ \ \ 
v_3 = \begin{bmatrix}
1 & -1           \\[0.3em]
1 & -1           \\[0.3em]

\end{bmatrix},\ \ \
v_4 = \begin{bmatrix}
-1 & 1           \\[0.3em]
-1 & 1            \\[0.3em]

\end{bmatrix};\] 
\[v_5 = \begin{bmatrix}
1 & 1           \\[0.3em]
-1 & -1           \\[0.3em]

\end{bmatrix},\ \ \
v_6 = \begin{bmatrix}
-1 & -1           \\[0.3em]
1 & 1            \\[0.3em]

\end{bmatrix};\ \ \ 
v_7 = \begin{bmatrix}
1 & -1           \\[0.3em]
-1 & 1           \\[0.3em]

\end{bmatrix},\ \ \
v_8 = \begin{bmatrix}
-1 & 1           \\[0.3em]
1 & -1            \\[0.3em]

\end{bmatrix}.\] 
Observe that the vertices \( v_1, v_3, v_5, v_7\) are orthogonal to each other  and of equal length \(2\) in terms of the Hilbert Schmidt inner product (which gives Euclidean structure).  The vertices \( v_2, v_4, v_6, v_8\) are their opposites. This implies that  the classical polytope \( {\bf LC}_{2,2}\) is congruent to \( 
2B_1^4\), the $4$-dimensional $\ell_1$-ball of radius $2$. In particular, we have $2^4=16$ facets (i.e., 3 dimensional faces).  
Each facet  of \( {\bf LC}_{2,2}\) is a convex hull of four vertices with neither two of them being opposite, 
 and  the normal vector to each facet is proportional to  the sum of its four determining vertices. For instance, the unit normal vector to the facets determined by \( v_1,v_3,v_5,v_7\) and \( v_1,v_3,v_5,v_8\) are respectively
\be E_{11}=\begin{bmatrix}
1 & 0           \\[0.3em]
0 & 0            \\[0.3em]
\end{bmatrix},\ \ \  {CHSH}_2 =
		\frac{1}{2}\begin{bmatrix}
1 & 1           \\[0.3em]
1 & -1            \\[0.3em]
\end{bmatrix}  \label{facets2by2}
\ee
The  facet determined by the latter normal vector  is referred to as  of CHSH-type; the corresponding determining inequality is called CHSH inequality.  It turns out (by exhausting all possibilities or by Fourier analysis argument) that all facets of \( {\bf LC}_{2,2}\) are among the above two types, that is, the normal vectors can be obtained by permuting rows or columns of either matrix (vector) in  (\ref{facets2by2}), or by multiplying them by $-1$. 
\(E_{11}\) is a representative of the first equivalent class of Bell inequalities whose members are \( \pm E_{ij}\), $i,j=1,2$, where $E_{ij}$ is the matrix with $1$ in the $ij$-th entry and $0$s elsewhere.  
From here on, we donote by \(E\) this class of facets (for all dimensions). They are of trivial-type in the sense that they cannot be violated by any quantum correlation, 
as follows easily from Lemma  \ref{descpnLCQC}) and the Cauchy-Schwarz inequality. 
For the CHSH-type  facets, the maximal violation,  well known as the Tsirelson's bound, is \( \sqrt{2}\) (see \cite{Tsirelson1985} and \cite{ABMB2017}, Proposition 11.11).

\begin{table}[h!]
\begin{center}
\caption{Classification of facets of ${\bf LC}_{2,2}$, which has $8$ vertices and $16$ facets.  The quantum value of a given facet is defined as the maximal violation of its determining Bell inequality on the set of quantum correlation matrices. 
\label{table1}}
\vspace{2ex}
\begin{tabular}{|c|c|c|c|}
	\hline 
	\rule[-1ex]{0pt}{2ex}  Facet type&   number of facets&
	\begin{tabular}{@{}c@{}}number of vertices \\ of each facet\end{tabular}    & quantum value \\ 
	\hline 
	\rule[-1ex]{0pt}{2ex}  $
	E$&  8& 4 & 1 \\ 
	\hline 
	\rule[-1ex]{0pt}{2ex} $
		CHSH$ &  8& 4 & $\sqrt{2}$ \\ 
	\hline 
\end{tabular} 
\vspace{1ex}
%
	\end{center}
\end{table}
Clearly all facets of same type have the same quantum value. 
As is well-known, \( K_G^{(2)}=\sqrt{2}\) (the Tsirelson bound)  is witnessed on the \(CHSH\) inequality. 
The principle, on which the argument is based, will be useful in what follows and so we state it here. 

\begin{lemma} \label{parallelogram} 
Let $v,w \in \mathcal{H}$ (an inner product space). Then 
\[
|v+w| +|v-w| \leq 2 (|v|^2+|w|^2)^{1/2}
\] 
with equality iff  $\bra v,w\ket = 0$. 
\end{lemma} 
{\bf Proof.}  We have 
\[
|v+w| +|v-w|  \leq \sqrt{2} (|v+w|^2 +|v-w|^2)^{1/2} = 2 (|v|^2+|w|^2)^{1/2}
\] 
by the Cauchy-Schwarz inequality and  the parallelogram identity.  The inequality becomes equality iff $|v+w| =|v-w|$, which is equivalent to 
 $\bra v,w\ket = 0$.   \qed
 
 \begin{prop}
 	The maximal quantum violation of the CHSH inequality is 
	\( K_G^{(2)}=\sqrt{2}\).
 \end{prop}
{\bf Proof.} 
 Let \(\varphi(A)=\Tr\left(\frac{1}{2}\begin{bmatrix}
 1 & 1           \\[0.3em]
 1 & -1            \\[0.3em]
 \end{bmatrix}A\right) \)  be the CHSH functional.  
 By combining Lemma \ref{descpnLCQC} and Remark \ref{reduction} we see that
 \[
 \max_{A\in QC_{2,2}} \varphi(A) = \frac 12 \max_{|x_1|=1,|x_2|=1} |x_1+x_2|+|x_1-x_2| 
 \]
 and it remains to apply Lemma \ref{parallelogram}.  The argument also characterizes 
 configurations $(x_i),(y_j)$, for which the maximal violation occurs:  we must have 
 $|x_1|=|x_2| = 1, \bra x_1,x_2\ket =0$ and $y_1= \frac{x_1+x_2}{|x_1+x_2|} =\frac1{\sqrt{2}}(x_1+x_2), 
 y_2= \frac1{\sqrt{2}}(x_1-x_2)$.  
 Retracing the proof of Tsirelson's Lemma \ref{descpnLCQC}, one may likewise find the corresponding 
operators \((X_i)\), \( (Y_j)\) and a state $\rho$ from the original definition of $QC_{2,2}$  
(cf. Definition \ref{def:qc}).  \qed

\vskip .25in
Concerning the classical polytope \( {\bf LC}_{3,3}\), the classification of the facets of \({\bf LC}_{3,3}\) (or more general, \( {\bf LC}_{3,n}\)) goes back to \cite{garg1983} (see Notes and Remarks to Section 11.2 in \cite{ABMB2017});  see also section 6.4 in \cite{avisetal4by4} , where a link to the theory of \emph{cut polytopes} 
is exploited. First,
it's routine to establish that \( {\bf LC}_{3,3}\) has 90 facets (though doing this manually would be rather tedious). 
It turns out that all of them can be obtained from 
\be E_{11}=\begin{bmatrix}
1 & 0  & 0         \\[0.3em]
0 & 0  & 0          \\[0.3em]
0 & 0 & 0          
\end{bmatrix} \ \hbox{ and }  \  {CHSH}_3 =
		\frac{1}{2}\begin{bmatrix}
1 & 1   & 0        \\[0.3em]
1 & -1   & 0         \\[0.3em]
0 & 0   & 0        
\end{bmatrix}  \nonumber
\ee
by permuting rows or columns and/or by multiplying them by $-1$. 
Indeed,  if \( \widehat{E}_{ij}\) is the subspace of \( 3\times 3\) matrices with the i-th row and j-th column  
being 0 and \(P_{ij}\)  
is the orthogonal projection onto \(\widehat{E}_{ij}\), then \( P_{{ij}}{\bf LC}_{3,3}={\bf LC}_{3,3} \cap \widehat{E}_{ij}\) 
can be identified with \({\bf LC}_{2,2}\).  Thus when \( \varphi\) is a Bell inequality determining a face of \({\bf LC}_{2,2}\), 
then $\varphi \circ P_{ij}$ determines a face of \( {\bf LC}_{3,3}\). 
So for each such projection, we obtain 8 faces of type \( E\) and 8 faces of type CHSH. 
Since there are  9 such projections, careful counting yields 18 faces equivalent to \(E_{11}\in M_3\)
and \( 8\times 9=72\) CHSH-type facets, which accounts for all 90 facets of  \({\bf LC}_{3,3}\).  
The results of the analysis are summarized in  Table \ref{table2}. 
\begin{table}[h!]
	\begin{center}
		\caption{Classification of facets of ${\bf LC}_{3,3}$, which has $32$ vertices and $90$ facets. \label{table2} } 
\vspace{2ex} 
\begin{tabular}{|c|c|c|c|}
	\hline 
	\rule[-1ex]{0pt}{2ex}  Facets type&   number of facets& \begin{tabular}{@{}c@{}}number of vertices \\ of each facet\end{tabular}    & quantum value \\ 
	\hline 
	\rule[-1ex]{0pt}{2ex}  $ E$
&  18& 16 & 1 \\ 
	\hline 
	\rule[-1ex]{0pt}{2ex} $CHSH$
&  72& 16 & $\sqrt{2}$ \\ 
	\hline 
\end{tabular} 
\vspace{1ex}

\end{center}
\end{table}

\noindent It follows that \( K_G^{(3)}\) again coincides the maximal quantum violation of the CHSH inequality, i.e.,  \( K_G^{(3)}=\sqrt{2}\). 

\vskip 0.3in
\section{$(4,4)$ Bipartite system} \label{section:4by4}

\subsection{Geometry of $(4,4)$ classical correlation polytope}

We now consider a bipartite quantum system with four dichotomic observables being measured  on each subsystem . 
In this case the classic correlation polytope is a convex polytope in \( \R^{16}\) with 128 vertices. Its facial structure can be completely described  \cite{qithannover, qitvienna} (verifed through Matlab package \cite{MPT3}, manual checking being rather infeasible).
 Besides the trivial type inequalities \(E\) and the \(CHSH\)-type inequalities

\[E_{11} = \begin{bmatrix}
1 & 0 & 0 & 0           \\[0.3em]
0 & 0 & 0 & 0           \\[0.3em]
0 & 0 & 0 & 0          \\[0.3em]
0 & 0 & 0 & 0            
\end{bmatrix},\ \ \
 {CHSH}_4= \frac{1}{2}\begin{bmatrix}
 1 & 1 & 0 & 0           \\[0.3em]
 1 & -1 & 0 & 0           \\[0.3em]
 0 & 0 & 0 & 0          \\[0.3em]
 0 & 0 & 0 & 0            
 \end{bmatrix},\] 
(and their relatives), 
only two new Bell correlation inequalities appear
( firstly proved in  \cite{avisetal4by4})
 \be 4_1 = \frac{1}{6}\begin{bmatrix}
 2 & 1 & -1 & 0           \\[0.3em]
 1 & -1 & 1 & 1           \\[0.3em]
 -1 & 1 & -1 & 1          \\[0.3em]
 0 & 1 & 1 & 0            
 \end{bmatrix},\ \ \ 
 4_2 = \frac{1}{10}\begin{bmatrix}
 1 & 1 & 2 & 2           \\[0.3em]
 1 & 2 & 1 & -2           \\[0.3em]
 2 & 1 & -2 & 1          \\[0.3em]
 2 & -2 & 1 & -1            
 \end{bmatrix} .\ee 
 Once all types have been identified, classifying the facets and identifying vertices belonging to them becomes routine.  
 The results are summarized in Table \ref{table3}.

\begin{table}[h!]
	\begin{center}
		\caption{Classification of facets of ${\bf LC}_{4,4}$, which has $128$ vertices and $27968$ facets.  
		\label{table3}} 
	\begin{tabular}{|c|c|c|c|}
		\hline 
		\rule[-1ex]{0pt}{2ex}  Facets type&   number of facets& \begin{tabular}{@{}c@{}}number of vertices \\ of each facet\end{tabular}   & quantum value \\ 
		\hline 
		\rule[-1ex]{0pt}{2ex}  
		$E$&  32& 64 & 1 \\ 
		\hline 
		\rule[-1ex]{0pt}{2ex} $
		CHSH$ &  288&  64 & \(\sqrt{2}\approx 1.414\)\\ 
		\hline 
		\rule[-1ex]{0pt}{2ex}  
		$4_1$&  18432&  24& $\frac{5}{3}\sqrt{2/3}\approx 1.3608$  \\ 
		\hline 
		\rule[-1ex]{0pt}{2ex} $
		4_2$ &  9216& 24 & \(\frac{2}{5}\sqrt{10+\sqrt{2}}\approx 1.3514\) \\ 
		\hline 
	\end{tabular} 
\vspace{1ex}

\end{center}
\end{table}

In the following we will find analytically the maximal quantum violations for  the inequalities 
\( 4_1\) and \( 4_2\).  It turns out that they are both smaller than $\sqrt{2}$. 
Accordingly,  \( K_G^{(4)}=\sqrt{2}\) again coincides with 
the maximal quantum violation of the CHSH inequality.

\subsection{Quantum violation for the \(4_1\) Bell correlation inequality}

By Remark \ref{reduction},  determining the 
maximal quantum violation  of \(4_1 = (m_{i,j})\) 
is equivalent to finding the maximum of \(E=\sum_{j=1}^4E_j\), where \(E_j=|\sum_{i=1}^46m_{i,j}x_i|\), 
over $|x_1|= |x_2|= |x_3|= |x_4|= 1$. 
Set \( u=x_2-x_3, v=x_2+x_3\) and denote \( t=\langle x_1,u\rangle, s^2=|u|^2\). Then 
     \begin{eqnarray*}
     	 E_1&=&|2x_1+u|=\sqrt{4+s^2-4t};\\
     	 E_2&=&|x_4+x_1-u|;\\
     	 E_3&=&|x_4-(x_1-u)|;\\
     	 E_4&=&|v|=\sqrt{4-s^2}.
     	 \end{eqnarray*}
     By Lemma \ref{parallelogram}, 
     \( E_2+E_3\le 2\sqrt{1+|x_1-u|^2}=2\sqrt{2+s^2+2t}\) with equality iff  \( x_4\perp x_1-u\).
     
     Next we find the maximum of \(E_1+E_2+E_3\) over the parameter \(t\in [-1,1]\). 
     Let \( \phi(t)=\phi_s(t) =\sqrt{4+s^2-4t}+2\sqrt{2+s^2+2t}\). 
   The first derivative test yields $\phi'(t)=0$ iff \( t=1/3\) (for all $s$)  
   and \( \phi''(1/3)<0\). Hence \(E_1+E_2+E_3\le  \phi(1/3) = 3\sqrt{8/3+s^2}\).
     
     Finally, we maximize \( E_1+E_2+E_3+E_4\) with respect to the parameter \(s \in [0,1]\). 
Collecting the bounds obtained thus far and applying Cauchy-Schwarz inequality we get 
     \begin{eqnarray*}
      E_1+E_2+E_3+E_4&\le &3\sqrt{8/3+s^2}+\sqrt{4-s^2} \\ 
      &\le & \sqrt{3^2+1^2}\sqrt{8/3+s^2+4-s^2}=10\sqrt{2/3}.
     \end{eqnarray*}
     The second inequality becomes the equality iff  
     \(\sqrt{8/3+s^2}=3\sqrt{4-s^2}\), that is, iff \(s=\sqrt{10/3}\). Therefore, as asserted,  
     \[ \sum_{j} \big| \sum_{i} m_{i,j} x_i \big|=\frac{E_1+E_2+E_3+E_4} 6\le \frac{5}{3}\sqrt{2/3}\approx 1.3608. \] 
     
To conclude, we need to verify that this bound is saturated, 
i.e., that the obtained extremal values of parameters are consistent. 
One way to achieve that is to retrace the argument and to identify an extremal configuration. 
The constraints are 
 \( \langle x_1,x_2+x_3\rangle =1/3, \langle x_2, x_3\rangle =-2/3\) and \( \langle x_1-x_2+x_3,x_4\rangle=0\) 
and they are satisfied, for example, by 
     \( x_1=(1,0,0,0), x_2=(0,1,0,0), x_3=(1/3, -2/3, 2/3,0), x_4=(0,0,0,1)\).

\subsection{Quantum violation for the \(4_2\) Bell correlation inequality}

We shall find  the maximum of \(F=\sum_{j=1}^4F_j\) where \(F_j=|\sum_{i=1}^410m_{i,j}x_i|\) 
and \(m_{i,j}\) is the \((i,j)\)-th entry of \(4_2\). 
By the Cauchy-Schwarz inequality, 
\be  \label{equality-4-2a}
 \sum_{j=1}^4 F_j\le 2 (\sum_{j=1}^4 F_j^2)^{1/2},\ee 
with equality if \( F_1=F_2=F_3=F_4\). 
We have 
\begin{eqnarray}
	 \left(\sum_{j=1}^4 F_j^2\right)^{1/2}&=&\left(40+2\langle x_1, x_2\rangle+2\langle x_1, x_3\rangle+2\langle x_2, x_4\rangle-2\langle x_3, x_4\rangle\right)^{1/2} \nonumber\\
	 &=& \left(40+2\langle x_1, x_2+x_3\rangle+2\langle x_2-x_3, x_4\rangle\right)^{1/2} \nonumber\\
	 &\le&(40+4\sqrt{2})^{1/2}. 
	 \label{equality-4-2b}
\end{eqnarray}
It follows that \( F\le 4\sqrt{10+\sqrt{2}}\) and, consequently, 
\be \sum_{j} \big|\sum_{i} m_{i,j} x_i\big| = F/10 \le \frac{2}{5}\sqrt{10+\sqrt{2}} \approx 1.3514 .\ee
We will show that this bound can be attained. 
First, the equality in \eqref{equality-4-2b} holds when 
\( x_2\perp x_3, x_1=(x_2+x_3)/\sqrt{2}\) and \( x_4=(x_2-x_3)/\sqrt{2}\), which is clearly a feasible configuration. 
Under these constraints, one checks directly that that \( F_1=F_2=F_3=F_4=\sqrt{10+\sqrt{2}}\). Indeed, 
\[ F_1=|x_1+x_2+2x_3+2x_4|=|(1+3/\sqrt{2})x_2+(2-1/\sqrt{2})x_3|=\sqrt{10+\sqrt{2}};\]
\[ F_2=|x_1+2x_2+x_3-2x_4|=|(2-1/\sqrt{2})x_2+(1+3/\sqrt{2})x_3|=\sqrt{10+\sqrt{2}};\]
\[ F_3=|2x_1+x_2-2x_3+x_4|=|(1+3/\sqrt{2})x_2-(2-1/\sqrt{2})x_3|=\sqrt{10+\sqrt{2}};\]
\[ F_4=|2x_1-2x_2+x_3-x_4|=|(1/\sqrt{2}-2)x_2+(1+3/\sqrt{2})x_3|=\sqrt{10+\sqrt{2}}.\] 
This means that we have then equality both in \eqref{equality-4-2a} and \eqref{equality-4-2b}, as needed.
An example of an extremal configuration is 
\( x_1=(\sqrt{2}/2, \sqrt{2}/2), x_2=(1,0), x_3=(0,1)\), and \(x_4=(\sqrt{2}/2, -\sqrt{2}/2)\).
\begin{rem} \label{mixed-luck} 
The reader may verify that extremal configurations $(x_i),(y_j)$ for the inequality  \(4_1\)  can be 
identified using the procedure from Remark \ref{lucky}, while the extremal configurations for 
\(4_2\) can not. 
\end{rem}

\vskip.25in

\noindent
{\bf Acknowledgement}
I would like to present special thanks to my supervisor, Prof. Stanis\l aw J. Szarek, 
for his continual support and assistance. 
Part of this research was performed during the Fall of 2017 while the author participated in 
the thematic program of \emph{Analysis in Quantum Information Theory} at 
the Institut Henri Poincar\'{e} in Paris, France. Thanks are due to IHP and to its staff for their hospitality, 
and to fellow participants for many inspiring interactions. 
I would also like to thank Dr. David Reeb for comments on the preliminary version.
The research of the author is supported in part by SJS's grants DMS-1600124 and DMS-1700168 
from the {National Science Foundation (U.S.A.)}.

\bibliographystyle{plain}
\bibliography{gro4by4}

\begin{thebibliography}{10}

\bibitem{ABMB2017}
G.~Aubrun and S.~Szarek.
\newblock {\em Alice and {B}ob meet {B}anach}, volume 223 of {\em Mathematical
  Surveys and Monographs}.
\newblock American Mathematical Society, Providence, RI, 2017.
\newblock The interface of asymptotic geometric analysis and quantum
  information theory.

\bibitem{avisetal4by4}
David Avis, Hiroshi Imai, and Tsuyoshi Ito.
\newblock On the relationship between convex bodies related to correlation
  experiments with dichotomic observables.
\newblock {\em Journal of Physics A: Mathematical and General}, 39(36):11283,
  2006.

\bibitem{CHSH1969}
J.~F. {Clauser}, M.~A. {Horne}, A.~{Shimony}, and R.~A. {Holt}.
\newblock {Proposed Experiment to Test {L}ocal {H}idden-{V}ariable {T}heories}.
\newblock {\em Physical Review Letters}, 23:880--884, October 1969.

\bibitem{paulsenetal2017}
K.~{Dykema}, V.~I. {Paulsen}, and J.~{Prakash}.
\newblock {Non-closure of the set of quantum correlations via graphs}.
\newblock {\em ArXiv e-prints 1709.05032}, September 2017.

\bibitem{EPR1935}
A.~Einstein, B.~Podolsky, and N.~Rosen.
\newblock Can quantum-mechanical description of physical reality be considered
  complete?
\newblock {\em Phys. Rev.}, 47(10), 1935.

\bibitem{qitvienna}
IQOQI~Institute for~quantum optics and quantum~information Vienna.
\newblock Open {Q}uantum {P}roblems.
\newblock December 2017.
\newblock \url{https://oqp.iqoqi.univie.ac.at/open-quantum-problems}.

\bibitem{garg1983}
Anupam Garg.
\newblock Detector error and einstein-podolsky-rosen correlations.
\newblock {\em Phys. Rev. D}, 28:785--790, Aug 1983.

\bibitem{grothendieck}
A.~Grothendieck.
\newblock R\'esum\'e de la th\'eorie m\'etrique des produits tensoriels
  topologiques.
\newblock {\em Resenhas}, 2(4):401--480, 1996.
\newblock Reprint of Bol. Soc. Mat. S\~ao Paulo {{\bf{8}}} (1953), 1--79 [
  MR0094682 (20 \#1194)].

\bibitem{MPT3}
M.~Herceg, M.~Kvasnica, C.N. Jones, and M.~Morari.
\newblock {Multi-Parametric Toolbox 3.0}.
\newblock In {\em Proc.~of the European Control Conference}, pages 502--510,
  Z\"urich, Switzerland, July 17--19 2013.
\newblock \url{http://control.ee.ethz.ch/~mpt}.

\bibitem{qithannover}
Quantum Information~Theory in~Hannover.
\newblock Open problems.
\newblock May 2016.
\newblock \url{https://qig.itp.uni-hannover.de/qiproblems/OpenProblems}.

\bibitem{Ki2017}
S.~Kinneswig.
\newblock Bell {I}nequalities and {G}rothendieck's {C}onstant, {B}achelor's
  thesis.
\newblock Leibniz Universit\"{a}t Hannover, June 2017.

\bibitem{slofstra2017}
W.~{Slofstra}.
\newblock {The set of quantum correlations is not closed}.
\newblock {\em ArXiv e-prints 1703.08618}, March 2017.

\bibitem{Tsirelson1985}
B.~S. Tsirelson.
\newblock Quantum analogues of {B}ell's inequalities. {T}he case of two
  spatially divided domains.
\newblock {\em Zap. Nauchn. Sem. Leningrad. Otdel. Mat. Inst. Steklov. (LOMI)},
  142:174--194, 200, 1985.
\newblock Problems of the theory of probability distributions, IX.

\end{thebibliography}

\vskip .5in
\noindent
{\small Department of Mathematics, Applied Mathematics and Statistics }\\
{\small Case Western Reserve University, Cleveland, Ohio 44106, U. S. A. }\\
{\small \tt bxl292@case.edu}\\

\end{document}